\documentclass[%
 reprint,
 amsmath,amssymb,
aps,
]{revtex4-2}
\usepackage{xcolor}
\usepackage{graphicx}
\usepackage{dcolumn}
\usepackage{bm}
\usepackage[normalem]{ulem}
\usepackage[export]{adjustbox}

\begin{document}

\newcommand{\keldysh}[1]{\left\langle \left\{T_K~ #1  \right\}\right\rangle}
\newcommand{\sq}[1]{\left[#1\right]}
\newcommand{\crl}[1]{\left\{#1\right\}}
\newcommand{\rnd}[1]{\left(#1\right)}
\newcommand{\ang}[1]{\left<#1\right>}
\newcommand{\red}[1]{\textcolor{red}{#1} }
\newcommand{\blue}[1]{\textcolor{blue}{#1} }
\newcommand{\delb}[1]{\textcolor{blue}{\sout{#1}}}
\newcommand{\del}[1]{\textcolor{red}{\sout{#1}}}

\title{Finite width of anyons changes their braiding signature}

\author{K. Iyer}
\email{kishore.iyer@cpt.univ-mrs.fr}
\author{F. Ronetti}
\author{B. Gr\'emaud}
\author{T. Martin}
\author{J. Rech}
\author{T. Jonckheere}
\affiliation{Aix Marseille Univ, Université de Toulon, CNRS, CPT, Marseille, France}%

\date{\today}

\begin{abstract}
Anyons are particles intermediate between fermions and bosons, characterized by a nontrivial exchange phase, yielding remarkable braiding statistics. Recent experiments have shown that anyonic braiding has observable consequences on edge transport in the fractional quantum Hall effect (FQHE). Here, we study transport signatures of anyonic braiding when the anyons have a finite width. We show that the width of the anyons, even extremely small, can have a tremendous impact on transport properties and braiding signatures. In particular, we find that taking the finite width into account allows us to explain recent experimental results on FQHE at filling factor $2/5$ [Ruelle et al., Phys. Rev. X \textbf{13}, 011031 (2023)]. Our work shows that the finite width of anyons crucially influences setups involving anyonic braiding, especially  when the exchange phase is larger than $\pi/2$.
\end{abstract}

\maketitle

Anyons are particles intermediate between bosons and fermions, characterized by fractional exchange statistics~\cite{Leinaas1977, wilczek82, arovas84}. These are proposed to occur in two spatial dimensions, and have found a solid experimental footing in the fractional quantum Hall effect (FQHE)~\cite{tsui82,laughlin83}. Fundamental interest and potential technological applications~\cite{nayak08} have fuelled intense activity leading to several theoretical proposals to detect anyonic statistics in FQHE~\cite{chamon97,safi01,vishveshwara03,bishara08,campagnano12, rosenow12, levkivskyi12, halperin11,lee19}. Only recently, experiments were able to detect anyonic statistics in the FQHE in the simplest filling fraction of $1/3$~\cite{bartolomei20,nakamura20,Lee2022,Kundu2023} as well as in the more complicated fraction of $2/5$~\cite{Ruelle2022, Glidic2022braiding,nakamura2023fabryperot}.

Transport experiments on FQHE edges have been successful in quantitatively extracting anyonic statistics by measuring current correlations. In contrast to spatial braiding as in the Fabry-Perot geometries, the physical mechanism at play here is time-domain braiding~\cite{Han2016}. In the latter,  anyons emitted from a source quantum point contact (QPC) form a braiding loop in time with anyon pairs excited at a QPC in FQHE. The braiding loop in time is due to the interference between two different time-ordered processes~\cite{Morel22,Lee2022nonAbelian}.  

Experiments are well captured by the theoretical formalism for $\nu = 1/3$ FQHE where the exchange phase is $\pi/3$~\cite{rosenow16}. However, experimental results at $\nu=2/5$, where theory predicts an exchange phase of $3\pi/5$, 
strongly differ from the theoretical predictions, even for a quantity as essential as the sign of the tunneling current. The multi-edge structure of $\nu = 2/5$ FQHE suggests the influence of inter-edge interactions or edge reconstruction, among others, as possible sources of the observed deviation from theoretical predictions. However, none of these candidates seem to explain the observed sign of tunneling current.

In existing calculations, the natural assumption is to neglect the  width of anyons. Indeed, while anyonic excitations always have a nonzero extension, it is deemed negligible as it is typically much smaller
than the average spacing between successive anyons, and the thermal  length of the system. In this Letter, we show that the finite width of 
the anyons, even small, 
significantly affects their braiding signatures, 
reflecting on the transport properties of the  system, in particular for composite fractions of the FQHE where the exchange phase is larger than $\pi/2$.
Beyond the specific system  we consider, our results show that the finite extension of anyons is an essential ingredient
for a correct description of setups involving anyonic braiding.

\emph{Time-domain braiding:--} We first review the basics of time-domain anyonic braiding~\cite{Han2016,lee19,Morel22,mora2022anyonic}. Consider the geometry of Fig.~\ref{fig:geometry_singleparticle},
showing a Hall bar in the FQHE with chiral edge states, equipped with a QPC operating in the weak tunneling regime, where tunneling
of quasiparticles between the edges can occur.
For the system at equilibrium, at all times, spontaneous processes
where a quasiparticle-quasihole (qp-qh) pair is created at the QPC do exist, leading to zero net current due to electron-hole symmetry. However,
 when a single quasiparticle (qp) on the upper edge impinges on the QPC, 
there is a nontrivial interference between the process where the qp-qh pair is created before, and the one where it is created after the arrival of the single qp.  The interference of these two processes leads to a braiding loop in time domain, with an overall coefficient $1-e^{2i\theta}$, where $\theta$
is the exchange phase between two qps. For fermions and bosons, the
overall contribution is zero, as $\theta = \pi \mbox{ or } 0$,
and the tunneling current is only due to direct
tunneling of the incoming qp through the QPC.
For anyons however, the cancellation is only partial due to their nontrivial braiding statistics, with for example $\theta = \pi/3$ for anyonic qps in  $\nu = 1/3$ FQHE.
It has been shown that this anyonic exchange dominantly contributes to physically measurable quantities at the QPC such as the average tunneling current and  current correlations~\cite{Morel22,Lee2022nonAbelian}. In this work, we consider the impact of the finite width of the incoming qp on the anyonic exchange process, and its consequences on the
tranport properties.

\emph{Model:--} We consider a Hall bar in the FQHE with chiral edge states, equipped with a QPC where tunneling between the edge states can occur (see Fig.~\ref{fig:geometry_singleparticle}). For simplicity, 
each edge is described as a  single mode Laughlin chiral Luttinger liquid. As shown in SM [\onlinecite{Note1}], this can capture 
the physics of complex composite fractions like $\nu=2/5$ by adapting the values of the parameters (see the discussion of Eq.~\eqref{eq:IT}). 
Up and down edge states are described in the total Hamiltonian as $H_{0u/d} = \frac{v_F}{4\pi}\int dx (\partial _x \phi^{u/d})^2 $, where $\phi^{u/d}$ denotes the bosonic mode on the upper/lower edge, $v_F > 0$ is the propagation velocity of the bosonic mode. The bosonic modes satisfy equal-time commutation relations $\left[\phi^{u/d}(x), \phi^{u/d}(y) \right] = \pm i\pi{Sign}(x-y)$. 
The edge hosts anyonic quasiparticles of charge $e^* = \nu e$, described by the operator $\psi \sim e^{i\sqrt{\nu}\phi}$. 
The QPC is placed in the weak backscattering regime, causing tunneling of anyonic qps between the two edges. The corresponding  tunneling Hamiltonian is $H_T(t) = \Gamma \left[ A(t) + A^\dagger(t) \right]$, where $A(t) = e^{i\sqrt{\nu}(\phi^u(0,t) -\phi^d(0,t))}$ and $\Gamma$ is the tunneling amplitude. Similarly, the current flowing from edge $u$ to $d$ is given by $I_T (t) = i e^*\Gamma \left[ A^\dagger(t) - A(t) \right]$.

\emph{Current due to a single qp:--} We first consider the current created by a single anyonic qp on the upper edge incident on the QPC. Crucially, this qp is taken to have a finite extension.
{Due to chirality of the edge, this spatial width leads to a finite temporal width
for the bosonic field at the position of the QPC. This can be modeled by adding a finite-width solitonic excitation on the bosonic edge modes~\cite{rosenow16, schiller2022anyon}
\begin{equation}
  \phi^{u} \longrightarrow \phi^{u} + 2 \sqrt{\lambda} \left[\tan^{-1}\left(\frac{t-t_0}{t_w} \right) + \frac{\pi}{2} \right]
  \label{eq:phiSingleqp}
\end{equation}
where $t_w$ denotes the temporal width of the qp,
whose center reaches the QPC at $t=t_0$
and $\pi \lambda$ is the phase due to the exchange of two qps. The current at the QPC due to this qp can be expressed, to leading order in $\Gamma$, using the Keldysh formalism as~\cite{martin05}
\begin{equation}
    \left<I_T(t)\right> = -\frac{i}{2}\int dt' \sum_{\eta\eta'} \eta'\keldysh{I_T(t^\eta) H_T(t'^{\eta'})}
    \label{keldysh_current}
\end{equation}
where $\eta, \eta'$ denote the Keldysh contour labels, and $T_K$ denotes time ordering on the Keldysh contour. It takes the form~\cite{Note1}
\begin{align}
    &\left<I_T(t)\right> = 2 i e^* \Gamma^2 \int_{-\infty}^{t}dt' \left( e^{2\delta\mathcal{G}(t-t')} - e^{2\delta\mathcal{G}(t'-t)} \right)
    \nonumber \\
    & \times \sin \Bigg\{2 \lambda \left[ \tan^{-1}\left(\frac{t-t_0}{t_w} \right)-\tan^{-1}\left(\frac{t'-t_0}{t_w} \right)\right]
    \Bigg\}
    \label{eq:IT}
\end{align}
where $\mathcal{G}(t) =  \left< T \phi(t)\phi(0)\right>$ is the bosonic Green's function
\begin{equation}
\mathcal{G}(t) = \ln \left(\frac{\sinh(i \pi k_B T\tau_0/\hbar)}{\sinh[\pi k_B T(t - i\tau_0)/\hbar]} \right) .
\end{equation}
Here $\tau_0$ is the short-time cutoff, $T$ is the temperature, and $\delta$ is the scaling dimension of the qp tunneling across the QPC. For the Laughlin case, one simply has $\lambda = \delta = \nu$.
For composite fractions, the values $\lambda$ and $\delta$ can be adjusted to describe the physics of the 
associated qp.
For example, for FQHE with filling factor $\nu = 2/5$, the $e/5$ qps can be addressed simply by taking $\lambda = \delta = 3/5$, and $e^* = e/5$, in Eq.~\eqref{eq:IT}.~\footnote{See Supplemental Material for details on the model with multiple bosonic edge modes \cite{shtanko14}, theory of the anyon collider, results for a different qp profile, and the relation between the braiding phase and the sign of the current.}\cite{Wen95}

The incoming qp wavepacket in Eq.~\eqref{eq:phiSingleqp} is assumed to have a Lorentzian profile~\cite{Note1}. Due to the finite width of the qp wavepacket, its braiding phase is smeared in time, and it contributes only partially at a given time. This can significantly affect the tunneling current as illustrated in Fig.~\ref{fig:geometry_singleparticle}(b). 
For $t_w \to 0$, the sine in Eq.~\eqref{eq:IT} reduces
to $\theta(t-t_0) \theta(t_0-t') \sin(2 \pi \lambda)$, where $\theta$ is the Heaviside step function, \and the current then reduces to a slowly decaying function of time, with a characteristic timescale set by the temperature, $t_\text{Th}=\hbar/(k_B T)$~\cite{jonckheere23}.

\begin{figure}
     \includegraphics[scale=0.35]{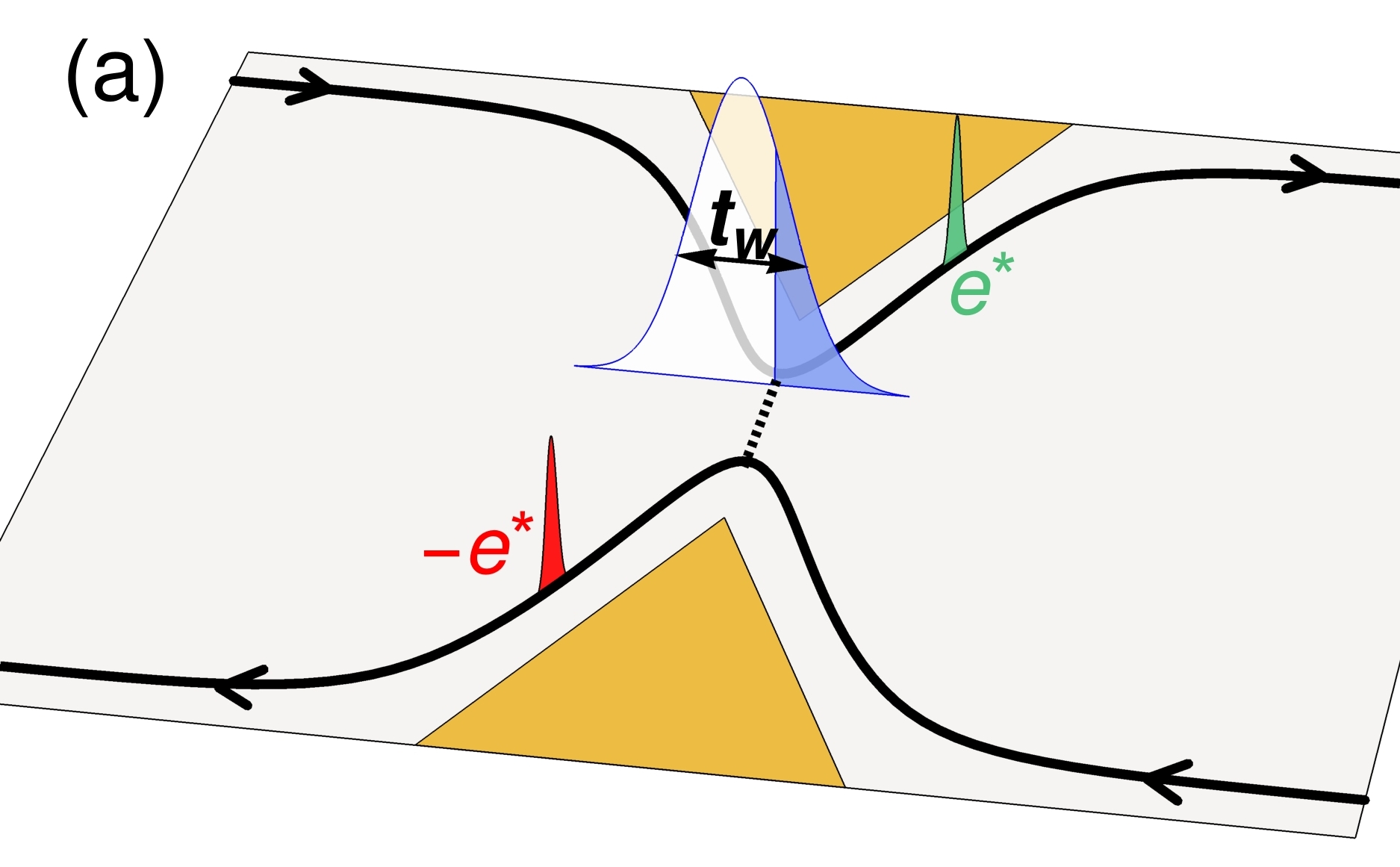}
    \\$\quad$\\
    \includegraphics[scale=0.5]{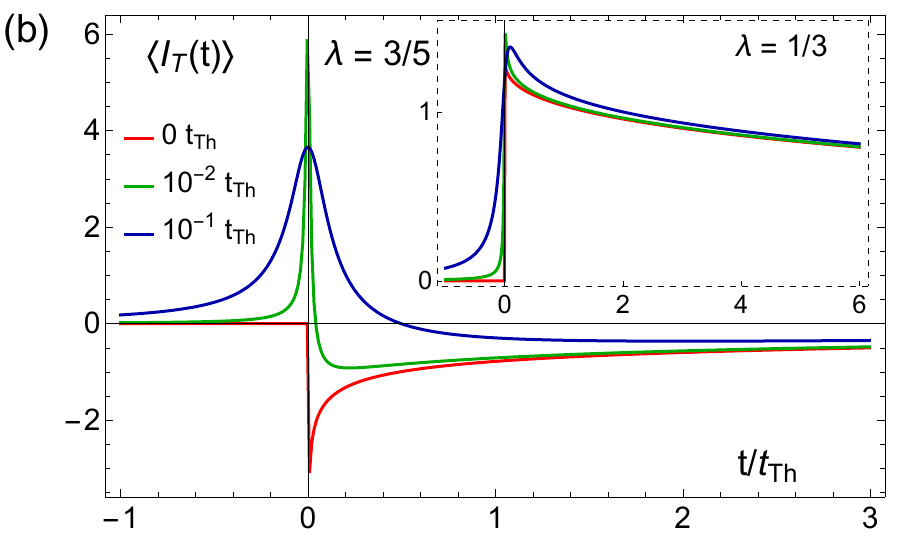}
    \caption{(a)  Idealized geometry of a FQHE bar with a QPC, showing an extended quasiparticle (qp) impinging on the QPC on the upper edge (in blue/white).
    A tunneling current through the QPC is created by
    the anyonic exchange between the qp and the qp-quasihole pairs (green-red) excited at the QPC.  Due to its finite temporal width ($t_w$), the qp 
    contributes only partially to braiding at time $t$ at the QPC, as indicated by the blue part of the qp. 
    (b) Tunneling current at the QPC, in units of $I_0 = \Gamma^2 T (\pi T \tau_0)^{2\delta-1}$, due to a single incoming qp as a function of time 
    (Eq.~\eqref{eq:IT}) for 
    $\lambda = \delta = 3/5$, and for different qp widths
    ($t_w = 0, 0.01, 0.1$ in units of $t_{Th} = \hbar/ k_B T$). The center of the qp hits the QPC at $t=0$. As the width is increased
     the current shifts from negative to positive values. Inset: same figure for $\lambda=\delta=1/3$, showing that the effect of the qp width is not significant for $\lambda<1/2$.}
    \label{fig:geometry_singleparticle}
\end{figure}

  We show the tunneling current as a function of time for $\lambda=3/5$ in Fig.~\ref{fig:geometry_singleparticle}(b) for different values of the width $t_{w}$ 
 of the incoming qp, and $\delta = 3/5$.
For $t_w=0$, the tunneling current is strictly negative, as $\sin(2 \pi \times 3/5) <0$. As soon as the qp has a finite width, the current becomes positive, on a time interval around $t_0$ which becomes wider as $t_w$ is increased. Importantly, even if $\delta$ is adjusted, the qualitative behavior of the currents remains largely unchanged, as long as $\lambda > 1/2$: changing $\delta$ modifies the power-law decrease of the Green functions
in Eq.(\ref{eq:IT}), which will mainly move the value of $t_w$
at which the current becomes negative, without changing the overall behavior of the current.
This behavior is in complete contrast to that at $\lambda=1/3$
(which corresponds to the usual Laughlin $\nu=1/3$ case).
 Here, the current due to a qp of zero width is positive. However, endowing the qp with a finite width does not significantly impact the current, as shown in the inset of
Fig.~\ref{fig:geometry_singleparticle},
for $\delta=1/3$. Again, the qualitative behaviour is robust against change of parameters, as long as $\lambda < 1/2$.

As shown in Fig.~\ref{fig:geometry_singleparticle},
the impact of a finite width is starkly different between $\lambda>1/2$ and $\lambda<1/2$. For $\lambda>1/2$, the smearing of the braiding phase due
to the finite width of the incoming anyon leads to an effective $\lambda$ smaller than 1/2
at small times, dramatically changing the sign of the current. On the other hand, for $\lambda<1/2$, the smearing
of the braiding phase only produces a small
quantitative effect, without qualitative change of the current. $\lambda=1/2$ acts as a threshold value: for $\lambda<1/2$ (resp. $\lambda>1/2$) the braiding at the QPC occurs dominantly between the incoming anyonic qp and the QPC qp (resp. qh), which explains
why the tunneling current changes sign when $\lambda$ crosses $1/2$ [\onlinecite{Note1}].

\emph{Poissonian stream of qps:-} 
We now consider the case of a Poissonian stream of qps incident on the QPC, leading to an average current $I_{u/d}$ on the upper/lower edge. This corresponds to the situation of Ref.~\onlinecite{bartolomei20} 
where upstream QPCs in the tunneling regime are used as the sources of the qp streams. 

In the single qp case, there were two time scales characterizing the system:  the thermal time $t_{Th} = \hbar/(k_B T)$ and, $t_w$, the temporal width of the qp. A stream of qps introduces another important time scale: the average temporal spacing between successive qps, given by the inverse of the incoming current on the QPC, $t_s = e^*/I_+$, where $I_\pm = I_u\pm I_d$. 

Normalizing all times with $t_s$, the stream of incoming qps on the QPC can be modelled by modifying the bosonic fields as
\begin{equation}
    \phi^{u/d} \longrightarrow \phi^{u/d} + 2 \sqrt{\lambda} \sum_k \left[ \tan^{-1}\left(\frac{(t-t^{u/d}_k)/t_s}{\tau_w} \right) + \frac{\pi}{2} \right]
\end{equation}
where $t^{u/d}_k$ denotes the time at which the $k-$th qp on the $u/d$ edge hits the QPC, and these times follow a Poissonian distribution. Moreover, we have defined the scaled width of qps, $\tau_w = t_w/t_s$. Using the expression of current in Eq.~\eqref{keldysh_current}, the average tunneling current at the QPC is~\cite{Note1}
\begin{align}
    \left\langle I_T\right\rangle&= -4 e^*\Gamma^2 \!\!\int_{0}^\infty \!\!\!\!\!\! dt~ \frac{\text{sin}\left[ x\, \text{Im} f(t,\lambda,\tau_w)\right]}{\text{exp}\left[ \text{Re} f(t,\lambda,\tau_w)\right]} 
     \mbox{Im}\left( e^{2\delta\mathcal{G}(t)}\right)
\label{eq:average_current}
\end{align}
where we have performed a Poisonnian average over the times $t_k^{u/d}$, and $x= I_-/I_+$ is the asymmetry of the incoming currents. $f(t,\lambda,\tau_w)$  is the phase accumulated as finite-width qps pass the QPC, and is given by
\begin{align}
    f(t,\lambda,\tau_w) = 
    \int_{-\infty}^{\infty}\!\!\! du \;  \Bigg\{ 1 -  
&\mbox{exp}\bigg(-2i \lambda\bigg[\tan^{-1}\left(\frac{t/t_s-u}{\tau_w}\right)
\nonumber \\
 &+\tan^{-1}\left(\frac{u}{\tau_w}\right)\bigg]\bigg) \Bigg\}   .
\label{eq:finite_width_phase}
\end{align}
Here, the integration over $u$ is essential as due to their finite width, the qps start affecting the QPC even before their center hits the QPC. We emphasize that Eq.~\eqref{eq:finite_width_phase} gives the phase accumulated due to braiding of the Poissonian stream of finite-width qps with qp-qh pairs at the QPC. This is readily seen by taking $t_w \to 0$, giving $f(t,\lambda,\tau_w) \to t/t_s(1-e^{-2i\pi\lambda})$, which is the phase due to a stream of zero-width qps~\cite{rosenow16}.

We plot the imaginary part of the finite-width phase $f(t,\lambda,\tau_w)$ as a function of time in Fig.~\ref{fig:Imag_finite_width_phase}, 
for $\lambda=3/5$ and different values of the
scaled width $\tau_w$. For $\tau_w=0$, the imaginary part of the phase is simply linear in $t$,
 with a negative slope equal to $\sin(2 \pi \lambda)$. As soon as the qps have a finite width, we find a dramatic change occurring close to $t=0$:
 $\mbox{Im} f(t,\lambda,\tau_w)$ is initially positive, and takes a finite amount of time (set by $\tau_w$) to become negative and eventually recover the same negative slope. This behavior is reminiscent of the behavior of the current due to a single qp shown in Fig.~\ref{fig:geometry_singleparticle}.  The inset of Fig.~\ref{fig:Imag_finite_width_phase} shows the imaginary part of the phase for $\lambda = 1/3$. There, we see a linear behavior with a positive slope for $\tau_w=0$, which is only marginally
affected when $\tau_w$ is non-zero. Note that the real part of the phase, which also enters Eq.~\eqref{eq:average_current}, mostly contributes to the integral at short times. However, it does not vary significantly with $\tau_w$ for any $\lambda$, and is thus not shown here.

\begin{figure}
    \centering
 \includegraphics[scale=0.5]{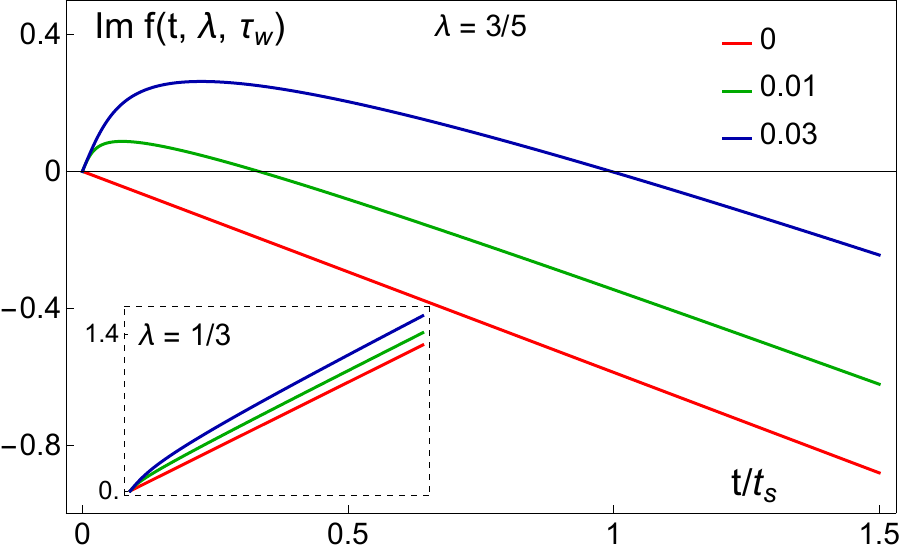}
\caption{Imaginary part of the finite-width phase as a function of time for $\lambda =3/5$, for scaled width $\tau_w = 0, 0.01, 0.03$. It is $\sin(2 \pi*3/5)(t/t_s)$ for $\tau_w=0$, but becomes positive close to $t=0$ when $\tau_w$ is finite.
Inset: same curves for $\lambda = 1/3$, showing that for $\lambda <1/2$ a finite width does not significantly change the phase.}
\label{fig:Imag_finite_width_phase}
\end{figure}

Hence, close to $t=0$, the effective phase experienced by the qps tunneling across the QPC is quite different from the full phase of the qp. As the 
rest of the integrand is peaked close to $t = 0$, a sizeable 
contribution to the tunneling current comes from the $t \simeq0$ region. We find that this results in an average positive tunneling current for $\lambda = 3/5$, as opposed to a negative average tunneling current seen for delta-width qps. This is shown in Fig.~\ref{fig:tunnelcurrent_pvalue}(a)
where the mean current $\langle I_T \rangle$ is plotted as a function
of the scaled width $\tau_w$ for different values of 
the asymmetry $x$ of the incoming currents. One readily sees that
$\langle I_T \rangle$ grows from negative to positive as $\tau_w$ is increased. For Lorentzian qp wavepackets, $\langle I_T \rangle$ changes sign at $\tau_w \sim 0.003$; for other shapes of finite width qps
(not shown), the qualitative behavior is the same. 

We now examine the consequence of the finite width of the incoming qps on the experimentally measured $P$ factor~\cite{Note1, rosenow16}, which is a generalized Fano factor for the cross-correlations of the output currents:  
\begin{equation}
    P(x=I_-/I_+) = \frac{\left<\delta I_u \delta I_d\right>}{~~e^*I_+\frac{\partial \left<I_T\right> }{\partial I_-} \Big|_{I_-=0}}
    \label{eq:pvalue}
\end{equation}
where $\left<\delta I_u \delta I_d\right>$ denotes the current cross-correlations. The $P$ factor was measured in recent experiments~\cite{bartolomei20,Ruelle2022,Glidic2022braiding} to extract anyonic statistics in FQHE. For $\nu = 1/3$ FQHE, the experiments find an excellent agreement with the theory of Ref.~\onlinecite{rosenow16} where the incoming qps are assumed to have zero width.
On the other hand, for $\nu =2/5$, the experiments are in strong disagreement with the theoretical calculations, as the $P$
factor is predicted to be positive and quite large (close to 6),
while the experiments measure negative values of the order of $-1$.
\begin{figure}
    \centering
    \includegraphics[scale=0.55]{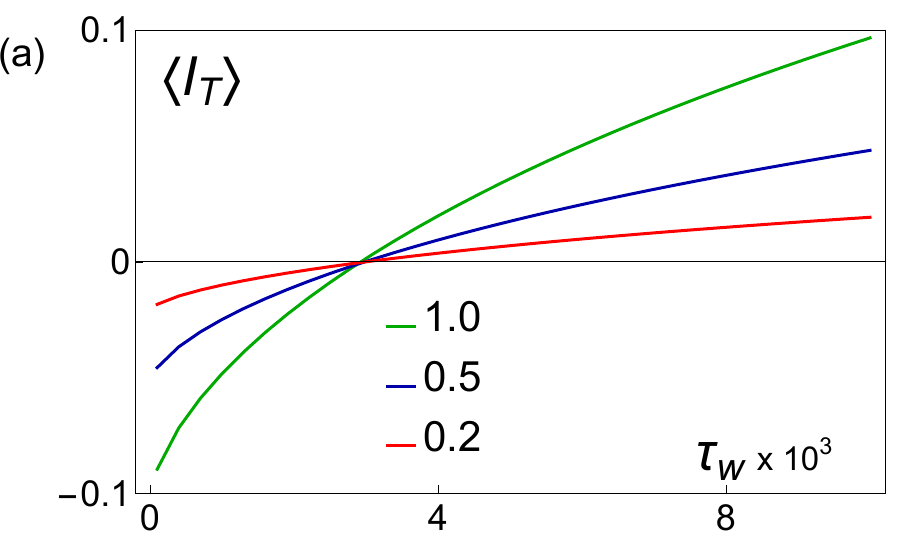}  \includegraphics[scale=0.55]{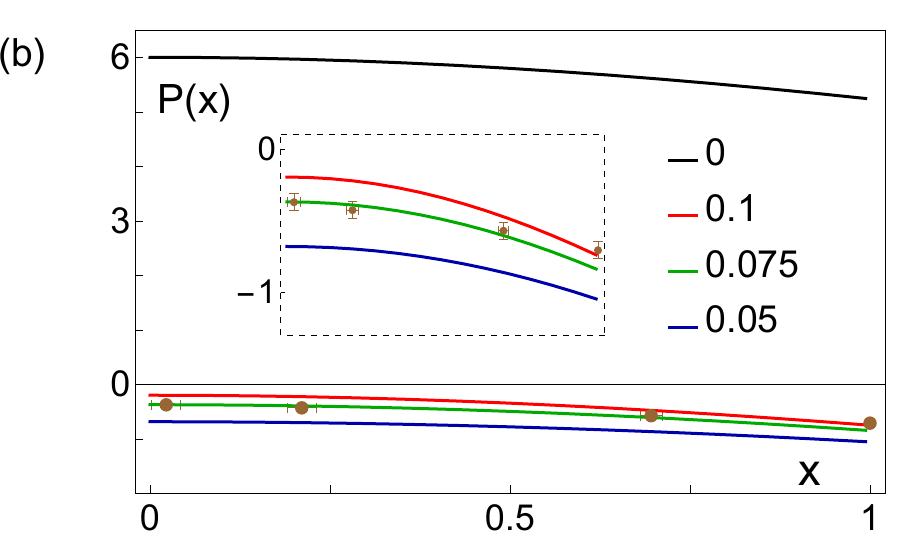}
  \caption{(a) Tunneling current $\langle I_T \rangle$ at the QPC due to a stream of Poissonian qps, as a function of scaled width $\tau_w$, for different values of current asymmetry: $x = 0.2, 0.5, 1$. $\langle I_T \rangle$ is negative for very
  small $\tau_w$, and becomes positive for $\tau_w \gtrsim 0.003$. (b) Fano factor $P(x)$ of Eq.~\eqref{eq:pvalue}  as a function of the current asymmetry $x$ for different values of scaled width: $\tau_w = 0, 0.05, 0.075, 0.1$. Experimental data from Ref.~\onlinecite{Ruelle2022} are shown as brown points with error bars. $P(x)$ is in contradiction even with the sign of the experimentally measured values for $\tau_w = 0$, but becomes compatible with the experimental data for $\tau_w \simeq 0.075$. Inset: zoom into the region close to the experimental data points. Curves on both panels were obtained in the regime of negligible temperature.}
\label{fig:tunnelcurrent_pvalue}
\end{figure}

We plot the $P$ factor for $\nu = 2/5$ FQHE ($\lambda = 3/5, \delta = 3/5$) in Fig.~\ref{fig:tunnelcurrent_pvalue}(b) for different values of the scaled width $\tau_w$. Experimental data points from Ref.~\onlinecite{Ruelle2022} are shown in brown. The black curve denotes the predictions of Ref.~\onlinecite{rosenow16} where $\tau_w =0$, and is in complete disagreement with the experimental data. With a non-zero $\tau_w$, the curves have now negative values, getting closer to the experimental data. The inset zooms into the region around the data points; we find that $\tau_w \sim 0.075$ agrees relatively well with the experiments.

The scaled width $\tau_w$ is proportional to the transparency $\mathcal{T}$ of the source QPCs. Indeed the applied voltage $V$ before a source QPC can be seen as a regular stream of incoming excitations whose spacing is equal to their width ($\sim 1/V$). As only a fraction $\mathcal{T}$  of these excitations is transmitted, it gives a
scaled width $\tau_w \sim \mathcal{T}$. The value $\tau_w \sim 0.075$ that we find 
to get a reasonable agreement with the experimental results for the $P$ factor
is thus compatible with the transparencies $\mathcal{T}$ used experimentally (typically $\mathcal{T} \sim 0.1$).
A detailed comparison with experimental results, including the effect of varying the transparency $\mathcal{T}$~\cite{Lee2022}, and considering different shapes for the excitations~\cite{Note1}, will be the subject of future work.

In conclusion, we have studied anyonic braiding signatures in FQHE edge transport accounting for the anyons' finite  width.
We have shown that the finite width of anyons decreases the effective braiding phase seen at the QPC. For a braiding phase $2 \pi\lambda > \pi$, a finite width can lead to an effective phase $< \pi$, changing the sign of the tunneling current. This allows us to quantitatively explain recent experiments on $\nu =2/5$ FQHE. In contrast, for a braiding phase $2 \pi\lambda < \pi$, the finite width only reduces this
phase further, causing no change in the sign of tunneling current. This explains the relative insensitivity of $\nu = 1/3$ FQHE to finite width of anyons. Our conclusions are robust against variation of the scaling dimension, which is typically non-universal. 

This work naturally leads to several possible extensions. Given the crucial impact of the finite width of incoming anyons, it might be interesting to study the consequences of the finite extent of the QPC~\cite{aranzana05,chevallier10,vannucci15}. 
The impact of Coulomb interaction inside the QPC could also be important for a complete description of the system. Finally, accounting for the finite width may be important in other architectures involving anyons including flying qubits~\cite{glattli20}, Fabry-Perot, and Mach-Zehnder interferometry~\cite{Carrega2021}, and even in different physical platforms hosting anyons such as spin liquids~\cite{Savary_2017,klocke21,klocke22,liu22}.

\acknowledgments
We thank G. F\`eve and M. Ruelle for useful discussions, and
for sharing their experimental data. We also thank M. Hashisaka and T. Kato for useful discussions.
This work was carried out in the framework of the project
``ANY-HALL'' (Grant ANR No ANR-21-CE30-0064-03).
It received support from the French government under the France 2030 investment plan, as part of the Initiative d'Excellence d'Aix-Marseille Université - A*MIDEX.
We acknowledge support from the institutes IPhU (AMX-19-IET-008) and AMUtech (AMX-19-IET-01X).
\textit{Note added in proof}:  after submission of this work, another work appeared on arXiv \cite{thamm2023finite}, which also studies the impact of finite width on anyon colliders, and finds
results consistent with ours.

\bibliography{finite_width_anyons_accepted_condmat.bbl}

\end{document}